\documentclass[12pt,psf,epsf]{article}
\usepackage[utf8]{inputenc}
\usepackage[english]{babel}
\pdfoutput=1

\usepackage[section]{placeins}

\usepackage{standalone}
\usepackage{ulem}
\usepackage{authblk}
\usepackage[centertags]{amsmath}
\usepackage{amssymb}
\usepackage[]{graphicx}
\usepackage{epsfig}
\usepackage{array}
\usepackage{amsthm}
\usepackage{latexsym}
\usepackage[mathcal]{euscript}
\usepackage{jheppubm}
\usepackage{hyperref}

\usepackage{psfrag}
\usepackage{float}
\usepackage{cancel}
\usepackage{mathrsfs}
\usepackage{amsfonts}
\usepackage{amsmath}
\usepackage{slashed}
\usepackage{bm}
\usepackage{color}
\usepackage{subcaption}

\newcommand{\be}{\begin{equation}}
\newcommand{\ee}{\end{equation}}
\newcommand{\bea}{\begin{eqnarray}}
\newcommand{\eea}{\end{eqnarray}}

\newcommand{\RNumb}[1]{\textbf{\uppercase\expandafter{\romannumeral #1\relax}}}

\usepackage{soul}
\usepackage{comment}
\usepackage{xcolor}

\usepackage{comment}

\begin{document}
\title{On the Equivalence Between the Schrodinger Equation in Quantum Mechanics and the Euler–Bernoulli Equation in Elasticity Theory}

\author{Igor Volovich}
\affiliation{Steklov Mathematical Institute, Russian Academy of Sciences, Gubkin Str. 8, 119991 Moscow, Russian Federation}

\emailAdd{volovich@mi-ras.ru}

\abstract{In this note, we show that the Schrodinger equation in quantum mechanics is mathematically equivalent to the Euler–Bernoulli equation for vibrating beams and plates in elasticity theory, with dependent initial data. 
Remarks are made on potential applications of this equivalence for symplectic and quantum computing, the two-slit experiment using vibrating beams and plates, and the \( p \)-adic Euler–Bernoulli equation.

}

\maketitle
\clearpage

\section{Introduction}

In 1750, Euler and Bernoulli proposed an equation to describe the vibrations of beams and plates in elasticity theory. The Euler–Bernoulli equation is given by \cite{Elasticity,Timoshenko}
\begin{equation}
\label{EB}
\ddot{u} + \Delta^2 u = 0,
\end{equation}
where $\Delta$ represents the Laplace operator. 

In 1926, Schrodinger discovered the wave equation that governs quantum mechanical particles. For a free particle, this equation is
\begin{equation}
\label{SE}
i \dot{\psi} = -\Delta \psi.
\end{equation}

Interestingly, these equations are fundamentally related. To derive the Euler–Bernoulli equation from the Schrodinger equation, we differentiate the Schrodinger equation with respect to time, yielding
\begin{equation}
\ddot{\psi} + \Delta^2 \psi = 0.
\end{equation}
Expressing $\psi$ as the sum of its real and imaginary parts yields two Euler–Bernoulli equations, each describing the vibration of a beam or plate. Thus, the Schrodinger equation corresponds to a system of two Euler–Bernoulli equations with dependent initial data. These initial data arise from two real functions derived from a single complex function that defines the initial condition for the Schrodinger equation.
\\

The paper is organized as follows. In Sect.~\ref{sect:CPSE} the Cauchy problem for the Schrodinger equation is discussed. In Sect.\ref{sect:CPEB}
the equivalence of a pair of the Euler–Bernoulli equations for vibrating beams and plates to the Schrodinger equation is proved. 
 In Sect.~\ref{ET-QQ}  remarks are made on potential applications of this equivalence for symplectic and quantum computing and  the two-slit experiment using vibrating beams and plates. In Sect.~\ref{sect:GEB} generalizations of the Euler--Bernoulli equation are considered.
In Sect.~\ref{sect:p-adic} p-adic version of the  Euler-Bernoulli equations is proposed.

\section{The Cauchy Problem for the Schrodinger Equation}\label{sect:CPSE}

Let us consider the following Cauchy problem for the Schrodinger equation:

\bea\label{2.1}
i \dot{\psi} &=& -\Delta \psi,\\
\label{2.2}
\psi(0,x)&=&\psi_0(x)
\eea
where $\psi = \psi (t,x)$, $x \in \mathbb{R}^d$, $t \in \mathbb{R}$, is a complex function, and
\begin{equation}
\Delta = \frac{\partial^2}{\partial x_1^2} + \dots + \frac{\partial^2}{\partial x_d^2}.
\end{equation}
We consider solutions $\psi \in C^\infty(\mathbb{R}^{d+1})$ with initial data $\psi_0 = \psi_0(x)$ belonging to the Schwartz space $\mathcal{S}(\mathbb{R}^d)$. The solution to the Cauchy problem \eqref{2.1}, \eqref{2.2} is given by the integral
\begin{equation}
\psi(t,x) = \int_{\mathbb{R}^d} \tilde{\psi}_0(k) \, e^{-ik^2 t - ik \cdot x} \, dk,
\end{equation}
where $\tilde{\psi}_0(k)$ is the Fourier transform of $\psi_0(x)$, with $k^2 = \sum_{i=1}^d k_i^2$ and $k \cdot x = \sum_{i=1}^d k_i x_i$.

Additionally, we have
\begin{equation}
\dot{\psi}(0,x) = i \Delta \psi_0(x).
\end{equation}

\subsection{Symplectic Form} 
We can rewrite the Schrodinger equation as a system of two equations for the real part $u$ and the imaginary part $v$ of the function $\psi$:
\begin{equation}\label{2.5}
\psi(x,t) = u(x,t) + i v(x,t),
\end{equation}
which leads to the system
\begin{equation}\label{2.6}
\dot{u} = -\Delta v, \quad \dot{v} = \Delta u.
\end{equation}
These equations represent Hamiltonian dynamics with the Hamiltonian
\begin{equation}\label{Hsym}
H_{\text{sym}} = \frac{1}{2} \int_{\mathbb{R}^d} \left[ u(-\Delta)u + v(-\Delta)v \right] \, dx.
\end{equation}

Note that
\begin{equation}
\partial_t u(0,x) = - \Delta v(0,x), \qquad \partial_t v(0,x) =  \Delta u(0,x).
\end{equation}

The Cauchy problem for the Schrodinger equation \eqref{2.1}, \eqref{2.2} is equivalent to the Cauchy problem for the system \eqref{2.6} with initial data
\begin{equation}
u(0,x) = u_0(x), \quad v(0,x) = v_0(x),
\end{equation}
where $u_0(x)$ and $v_0(x)$ are related to the initial data for the Schrodinger equation by
\begin{equation}
\psi_0(x) = u_0(x) + i v_0(x).
\end{equation}

The Fourier form of the system \eqref{2.6} is
\begin{subequations}
\begin{align}
\partial_t \tilde{u}(t,k) &= k^2 \tilde{v}(t,k), \\
\partial_t \tilde{v}(t,k) &= -k^2 \tilde{u}(t,k),
\end{align}
\end{subequations}
with initial data
\begin{equation}
\tilde{u}(0,k) = \tilde{u}_0(k), \quad \tilde{v}(0,k) = \tilde{v}_0(k).
\end{equation}

The Schrodinger equation forms a Petrovsky well-posed system of equations \cite{GelfandShilov:1967}.

\section{The Cauchy Problem for the Euler–Bernoulli Equation}\label{sect:CPEB}

Differentiating the Schrodinger equation \eqref{2.1} with respect to $t$, we obtain
\begin{equation}\label{2.1n}
\ddot{\psi} + \Delta^2 \psi = 0.
\end{equation}
Thus, the real and imaginary parts of $\psi = u + iv$ satisfy the Euler–Bernoulli equation
\begin{equation}
\label{assd}
\ddot{u} + \Delta^2 u = 0
\end{equation}
with initial data
\begin{equation}
\label{iccc}
u(0,x) = u_0(x), \qquad \dot{u}(0,x) = -\Delta v_0(x),
\end{equation}
and
\begin{equation}
\label{assdu}
\ddot{v} + \Delta^2 v = 0
\end{equation}
with initial data
\begin{equation}\label{icccm}
v(0,x) = v_0(x), \qquad \dot{v}(0,x) = \Delta u_0(x).
\end{equation}

The Euler–Bernoulli equation is a Petrovsky well-posed system of equations \cite{Sobolev}. In \cite{CorZuc}, solutions to the Euler–Bernoulli equations with potential are considered. The Cauchy problems for the Euler–Bernoulli equations \eqref{assd}, \eqref{iccc} and \eqref{assdu}, \eqref{icccm} in Fourier transform form are given by
\begin{equation}
\label{3.6}     
\tilde{u}(t,k) = \tilde{u}_0(k) \cos(k^2 t) + \tilde{v}_0(k) \sin(k^2 t),
\end{equation}
and
\begin{equation}
\label{3.7}
\tilde{v}(t,k) = \tilde{v}_0(k) \cos(k^2 t) - \tilde{u}_0(k) \sin(k^2 t),
\end{equation}
respectively.

We observe that the solutions \eqref{3.6} and \eqref{3.7} of the Cauchy problems for the Euler–Bernoulli equations satisfy
\begin{align}
\label{3.8}
\frac{\partial}{\partial t} \tilde{u}(t,k) &= k^2 \tilde{v}(t,k), \\
\label{3.9}
\frac{\partial}{\partial t} \tilde{v}(t,k) &= -k^2 \tilde{u}(t,k),
\end{align}
with initial data $\tilde{u}_0$ and $\tilde{v}_0$, respectively. This system is equivalent to the Hamiltonian equations \eqref{2.6}, which are in turn equivalent to the Schrodinger equation \eqref{2.1}.
\\

We have thus proven the following:
\\

\textbf{Theorem.} \textit{The function $\psi$ is a solution of the Schrodinger equation \eqref{2.1} with initial data $\psi_0(x) = u_0(x) + i v_0(x)$ belonging to  the Schwartz space $\mathcal{S}(\mathbb{R}^d)$  if and only if $u(t,x)$ and $v(t,x)$ satisfy the Euler–Bernoulli equations with initial data $u(0,x) = u_0(x)$, $\dot{u}(0,x) = -\Delta v_0(x)$, $v(0,x) = v_0(x)$, and $\dot{v}(0,x) = \Delta u_0(x)$. This solution is also equivalent to the solution of the system \eqref{2.6} with initial data $u_0$ and $v_0$.}

\section{Elasticity theory and Quantum Mechanics}
\label{ET-QQ}
In elasticity theory, Hooke's law is expressed as
\begin{equation}
u_{ik} = \lambda_{iklm} \sigma_{lm},
\end{equation}
where $u_{ik}$ is the deformation tensor, defined as $u_{ik} = \frac{1}{2}(d_i \partial_i u_k + d_k \partial_k u_i)$, and $\sigma_{lm}$ is the stress tensor. Using elasticity constants, the Euler–Bernoulli equation becomes
\begin{equation}
\mu \ddot{u} + TI \Delta^2 u = 0,
\end{equation}
where $\mu$ and $TI$ are material constants related to elasticity.

Comparing this with the Schrodinger equation 
\begin{equation}
i \hbar \dot{\psi} = -\frac{\hbar^2}{2m} \Delta \psi,
\end{equation}
we find the relationship
\begin{equation}
\frac{\hbar^2}{4m^2} = \frac{FI}{\mu},
\end{equation}
where $\hbar$ is the Planck constant, $m$ is the particle mass, and $F$ and $I$ are elasticity constants.

Furthermore, there is a correspondence between the natural frequencies of vibrating beams and plates and the eigenvalues of the quantum mechanical Hamiltonian. To derive the natural frequencies in elasticity theory, we consider the Euler–Bernoulli equation on a finite interval with specific boundary conditions. Seeking solutions in the form
\begin{equation}
u = \text{Re}(u(x) \exp(-i\omega t)),
\end{equation}
leads to an eigenvalue problem for $\omega$, analogous to the eigenvalue problem for the quantum mechanical Hamiltonian.
\\

It is shown in \cite{Volovich:2024nzw} that, in a finite-dimensional Hilbert space, the Schrödinger equation with an arbitrary Hermitian Hamiltonian is equivalent to a classical Hamiltonian symplectic orthogonal system. This insight is used to propose a symplectic computer, which could potentially surpass quantum computers in computational power. The Schrödinger equation, reformulated as a system of two equations for the real and imaginary parts of the wave function, governs the symplectic evolution of a classical mechanical system. This symplectic framework may enhance the capabilities of both quantum and symplectic computers \cite{Volovich:2024nzw}. 

Furthermore, it can be speculated that a system of vibrating beams or plates might serve as a toy model for quantum and symplectic computers. Interestingly, the two-slit experiment, traditionally described by the Schrödinger equation, can now be interpreted in terms of elasticity theory. 
This allows us to associate the quantum mechanical problem of scattering on a plane with coordinates \(x\) and \(y\) and a wave function \(\psi(x,y)\) with boundary conditions \(\psi(t,0,y) = 0\) for all \(y\), except for two holes, with the corresponding problem determined by the Euler–Bernoulli equation for  vibrating plates.

The Hamiltonian structure associated with the Schrodinger equation is discussed in \cite{VKS}. Mathematical foundations of the path integral approach to quantum mechanics are considered in  \cite{Shiryaev}.  The symplectic geometry of an infinite-dimensional phase space is examined in \cite{Khrennikov}. Infinite-dimensional generalizations of the Jacobi theorem is presented in \cite{SakVol}.
\section{Generalizations of the Euler--Bernoulli Equation}\label{sect:GEB}

In this section, we consider generalizations of the Euler--Bernoulli equation derived from the Schrodinger equation.

\subsection{Generalized Euler--Bernoulli Equation with Potential}

We start from the Schrodinger equation with a potential:
\bea
\label{cVm} i\dot{\psi} &=& H\psi, \\
H &=& -\Delta + V, \\
\label{idH} \psi(0,x) &=& \psi_0(x).
\eea
Differentiating \eqref{cVm} with respect to time, we obtain a generalization of the Euler--Bernoulli equation \eqref{2.1n} in the form
\bea
\ddot{\psi} + H^2 \psi = 0.
\eea
For $\psi = u + i v$, this leads to
\bea 
\label{uddH}
\ddot{u} + H^2 u &=& 0, \\
\label{vddH}
\ddot{v} + H^2 v &=& 0,
\eea
with initial data
\bea 
\label{iucH}
u(0,x) &=& u_0(x), \quad \dot{u}(0,x) = H v_0(x), \\
\label{ivH}
v(0,x) &=& v_0(x), \quad \dot{v}(0,x) = H u_0(x).
\eea
The Schrodinger equation \eqref{cVm} can also be expressed as 
\bea
\label{Suv}
\dot{u} = H v, \quad \dot{v} = -H u.
\eea
If $H$ is a self-adjoint operator, then the solution of the system \eqref{uddH} and \eqref{vddH} with initial data \eqref{iucH} and \eqref{ivH} is
\bea
u(t,x) &=& \cos(Ht) u_0(x) + \sin(Ht) v_0(x), \\
v(t,x) &=& \cos(Ht) v_0(x) - \sin(Ht) u_0(x).
\eea
Thus, the Schrodinger equation \eqref{cVm} with initial data \eqref{idH} is equivalent to a pair of generalized Euler--Bernoulli equations with interdependent initial conditions.

To conclude this subsection, we explicitly write the generalized Euler--Bernoulli equation with potential as
\be 
\label{eq:EB}
\ddot{u} + \Delta^2 u - \Delta (v u) - v \Delta u + v^2 u = 0.
\ee

\subsection{Elasticity Theory and Quantum Mechanics in Curved Space  with potential}
Let us consider the free particle in a curve space. In this case the Schrodinger equation has the form
\bea\label{csm} i\dot \psi&=&-\Delta_g \psi,\eea
where 
\bea
\label{cdeltam}\Delta_g&=&\frac{1}{\sqrt{g}}\, \partial _i (\sqrt{g}\,g^{ik} \partial _k )
\eea
is  the Beltrami-Laplace operator,   $g^{ik}=g^{ik}(x)$ is a positive defined metric in ${\mathbb R}^2$, $g=\det (g_{ik})$.

Differentiating \eqref{csm} on time we get the same Euler-Bernoulli equation \eqref{EB} with $\Delta $ given by \eqref{cdeltam}.

\section{P-adic Euler-Bernoulli equations}\label{sect:p-adic}
In this section, we consider a \( p \)-adic analogue of the Euler--Bernoulli equation. For background on \( p \)-adic mathematical physics, see 
\cite{IV-padic,Volbook,Dragovich:2017kge}. p-adic two-slit experiment is considered in \cite{Zuniga-Galindo:2023ost}.

We consider the following Cauchy problem for the Schrodinger equation:
\bea
i \dot{\psi} &=& D^\alpha \psi, \\
\psi(0,x) &=& \psi_0(x),
\eea
where \(\psi = \psi (t,x)\), with \( x \in \mathbb{Q}_p^d \), \( t \in \mathbb{R} \) is a complex function, and \( D^\alpha \) denotes the Vladimirov operator \cite{Volbook}. The initial data $\psi_0$ belong to the test functions.

Differentiating the Schrodinger equation \eqref{2.1} with respect to \( t \), we obtain
\bea
\ddot{\psi} + D^{2\alpha} \psi = 0,
\eea
which we call the complex \( p \)-adic Euler--Bernoulli equation. Thus, the real and imaginary parts of \(\psi = u + iv\) satisfy the \( p \)-adic Euler--Bernoulli equations:
\be 
\label{assdp}
\ddot{u} + D^{2\alpha} u = 0,
\ee
with initial data
\bea 
\label{icccp}
u(0,x) &=& u_0(x), \quad \dot{u}(0,x) = D^{\alpha} v_0(x),
\eea
and
\be 
\label{assdup}
\ddot{v} + D^{2\alpha} v = 0,
\ee
with initial data
\bea
\label{icccmp}
v(0,x) &=& v_0(x), \quad \dot{v}(0,x) = -D^{\alpha} u_0(x).
\eea
  The solutions of the Cauchy problem for the p-adic Euler--Bernoulli equations \eqref{assdp}, \eqref{icccp} and \eqref{assdup}, \eqref{icccmp} in the p-adic Fourier transform form are given by
\bea
\label{3.6p}
\tilde{u}(t,k) &=& \tilde{u}_0(k) \cos(|k|_p^\alpha t) + \tilde{v}_0(k) \sin(|k|_p^\alpha t), \\
\label{3.7p}
\tilde{v}(t,k) &=& \tilde{v}_0(k) \cos(|k|_p^\alpha t) - \tilde{u}_0(k) \sin(|k|_p^\alpha t),
\eea
respectively.

We observe that the solutions \eqref{3.6p} and \eqref{3.7p} of the Cauchy problems for the p-adic Euler--Bernoulli equations satisfy
\bea
\label{3.8p}
\frac{\partial}{\partial t} \tilde{u}(t,k) &=& |k|_p^\alpha \tilde{v}(t,k), \\
\label{3.9p}
\frac{\partial}{\partial t} \tilde{v}(t,k) &=& -|k|_p^\alpha \tilde{u}(t,k).
\eea
These equations represent Hamiltonian dynamics with the Hamiltonian
\begin{equation}\label{Hsym}
H_{\text{sym}} = \frac{1}{2} \int_{\mathbb{Q}^d} \left( u \,D^\alpha\,u + v \,D^\alpha\,v \right) \, dx.
\end{equation}

Therefore, the equivalence between the \( p \)-adic Schrödinger equation and the \( p \)-adic Euler–Bernoulli equation is established.

\section{Discussion and Conclusion}

In this note, we demonstrate that the Schrodinger equation is equivalent to a pair of Euler--Bernoulli equations describing two vibrating beams or plates, with dependent initial data. These initial data arise from two real functions obtained from a single complex function that defines the initial condition for the Schrodinger equation. 
\\

The Schrodinger equation, expressed as a system of two equations for the real and imaginary parts of the wave function, defines the symplectic evolution of a classical mechanical system. This symplectic formulation can be used to enhance the capabilities of quantum and symplectic computers \cite{Volovich:2024nzw}.
\\

It can also be speculated that a system of vibrating beams or plates could serve as a toy model for quantum and symplectic computers.
It is interesting that the two-slit experiment usually described by the 
 Schrodinger equation can be now described in term of elasticity theory.
\\

A geometric theory of defects in solids, based on classical elasticity theory and the Riemann-Cartan geometry, is suggested in \cite{Katanaev:1992kh}. Since this paper establishes an equivalence between the Euler--Bernoulli and Schrodinger equations, it seems promising to develop a theory of defects grounded by the Schrodinger equation.

\section*{Acknowledgement}

I am grateful to I.~Aref'eva, A.~Gushchin, M.~Katanaev, A.~Trushechkin, V.~Sakbaev, D.~Stepanenko, and V.~Zharinov for useful  discussions and support. 
This work is supported by the NSF Russian Science Foundation (24-11-00039, Steklov Mathematical Institute).

\end{document}